\documentclass[prl,aps,twocolumn,showpacs]{revtex4}
\usepackage{graphicx}
\begin{document}
\title{Optical analogue of spontaneous symmetry breaking induced by tachyon condensation in amplifying plasmonic arrays}
\author{A. Marini$^1$, Tr. X. Tran$^{1,2}$, S. Roy$^{1,3}$, S. Longhi$^4$ and F. Biancalana$^{1,5}$}
\email{Andrea.Marini@mpl.mpg.de}
\affiliation{$^1$Max Planck Institute for the Science of Light, Guenther-Scharowsky-Stra\ss e 1, 91058 Erlangen, Germany}
\affiliation{$^2$Department of Physics, Le Quy Don University, 236 Hoang Quoc Viet Street, 10000 Hanoi, Vietnam}
\affiliation{$^3$Department of Physics and Meteorology, Indian Institute of Technology, Kharagpur-721302, India}
\affiliation{$^4$Dipartimento di Fisica, Politecnico di Milano, and IFN-CNR, Piazza L. da Vinci 32, I-20133 Milano, Italy}
\affiliation{$^5$School of Engineering \& Physical Sciences, Heriot-Watt University, Edinburgh, EH144AS, United Kingdom}
\date{\today}
\begin{abstract}
We study analytically and numerically an optical analogue of tachyon condensation in amplifying plasmonic arrays. 
Optical propagation is modeled through coupled-mode equations, which in the continuous limit can be converted into a 
nonlinear one-dimensional Dirac-like equation for fermionic particles with imaginary mass, i.e. fermionic tachyons. We 
demonstrate that the vacuum state is unstable and acquires an expectation value with broken chiral symmetry, 
corresponding to the homogeneous nonlinear stationary solution of the system. The quantum field theory analogue of this 
process is the condensation of unstable fermionic tachyons into massive particles. This paves the way for using amplifying 
plasmonic arrays as a classical laboratory for spontaneous symmetry breaking effects in quantum field theory.

\end{abstract}
\maketitle

\paragraph{Introduction --} Photonic crystals and their one-dimensional realizations -- waveguide arrays 
(WAs) -- have been extensively studied in order to mimic the non-relativistic dynamics of quantum particles in 
periodic potentials \cite{ChristoNat2003,LonghiLasPhotRev2009,GaranovichPhysRep2012}. In this respect, WAs constitute a 
useful classical laboratory for simulating quantum effects and can be used either to analyze well-known fundamental mechanisms 
like Bloch oscillations \cite{PertschPRL1999}, Zener tunneling \cite{GhulinyanPRL2005,TrompeterPRL2006}, optical dynamical 
localization \cite{LonghiPRL2006}, and Anderson localization in disordered lattices \cite{LahiniPRL2008},
or even possibly uncover novel quantum effects. The thorough correspondence between the Schr\"odinger equation for
the quantum wavefunction and the paraxial equation for the optical field is the key that makes it possible to 
establish a precise quantum-optical analogy. Similarly, it is possible to mimic relativistic phenomena of quantum 
field theories in binary waveguide arrays (BWAs), since optical propagation in the continuous limit is governed by 
a (1+1)D Dirac equation \cite{ZeunerPRL2012}. Several mechanisms such as Klein tunneling \cite{LonghiPRB2010},
Zitterbewegung \cite{DreisowPRL2010}, Klein paradox \cite{DreisowEPL2012}, and fermion pair production
\cite{LonghiAPB2011} have been observed in BWAs. Analytical soliton solutions of the discrete coupled-mode equations
(CMEs) for a BWA, constituting the optical analogue of the (1+1)D nonlinear relativistic Dirac equation, have been 
recently reported \cite{TranPRL2013}. Although there is no evidence of fundamental 
nonlinearities in quantum field theory (QFT), the nonlinear Dirac equation has constituted a matter of study since long time
and it has been used as an effective theory in atomic \cite{IonescuPRA1988}, nuclear and 
gravitational physics \cite{ZeccaJTP2002} and in the study of ultracold atoms 
\cite{HaddadEPL2011}. An intriguing mechanism arising in quantum field theories is represented by 
{\it tachyon condensation} \cite{FeinbergPhysRev1967}. This is a process in particle physics where the system lowers 
its energy by spontaneously generating particles. The tachyonic field with complex mass is unstable and acquires a 
vacuum expectation value reaching the minimum of the potential energy and getting a non-negative squared mass. 
This mechanism is intimately related to the process of {\em spontaneous symmetry breaking}, i.e. the spontaneous collapse 
of a system into solutions that violate one or more symmetries of the governing equation, which in other contexts is 
responsible for the existence of Higgs bosons \cite{HiggsPRL1964}, Nambu-Goldstone bosons 
\cite{NambuPhysRev1960,GoldstonePhysRev1962} and fermions \cite{SalamPhysLettB1974}. 

Motivated by the importance of using BWAs as a classical laboratory for the study of QFT phenomena, 
in this Letter we theoretically investigate optical propagation in amplifying plasmonic arrays with alternate couplings, 
which in the continuous limit are governed by a nonlinear Dirac-like equation with imaginary mass. We find that the vacuum 
state is unstable and acquires an expectation value with {\em broken chiral symmetry} corresponding to the dissipative 
nonlinear stationary mode. We also study modulational instability, finding the conditions where the new vacuum is stable
and unstable due to the presence of topological defects, i.e. dissipative solitons. This paves the way for using BWAs 
to simulate tachyon condensation and spontaneous symmetry breaking mechanisms arising in QFT.

\begin{figure}[t]
\centering
\begin{center}
\includegraphics[width=0.45\textwidth]{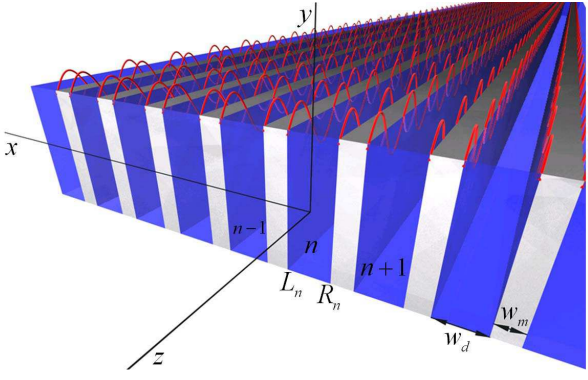}
\caption{Illustrative sketch of the structure analyzed in this work: a metal-dielectric stack supporting SPPs at every 
$z-y$ interface. The dielectric media (blue slabs) embed externally pumped active inclusions, which 
amplify SPPs (red sinusoidal curves) propagating along the $z$-direction. Every $n$th dielectric slab 
of width $w_d$, adjacent to a metallic stripe of thickness $w_m$, supports SPPs at the left and right interfaces with 
optical amplitudes $L_n$, $R_n$.
}
\label{Fig1}
\end{center}
\end{figure}

\paragraph{Model --} In the following we consider an amplifying plasmonic array -- a metal-dielectric
stack -- sketched in Fig. \ref{Fig1}. Surface plasmon polaritons (SPPs) propagating at every metal-dielectric interface 
are weakly coupled to nearest neighbours through alternating positive and negative couplings \cite{MariniOL2010}.  
This condition can also be achieved in BWAs either through a Bragg structure with a low-index defect 
\cite{EfremidisPRA2010} or through waveguides with propagation constants that vary periodically along the propagation 
direction \cite{ZeunerPRL2012,SzameitPRL2009}. Amplification schemes with SPPs have been intensively studied and also demonstrated 
experimentally \cite{MariniOL2009,BeriniNatPhot2011}. Gain is provided by externally pumped active inclusions embedded 
in the dielectric layers that can be modeled as two-level atoms. For continuous monochromatic waves oscillating with 
angular frequency $\omega$, the complex susceptibility $\epsilon_d$ of the pumped dielectric media is inherently 
nonlinear \cite{MariniPRA2010}: $\epsilon_d = \epsilon_b + \alpha (\delta - i)/(1 + \delta^2 + |{\bf E}/E_S|^2)$, where 
$\epsilon_b$ is the linear susceptibility of the hosting medium, $\alpha$ is the dimensionless gain rescaled to 
$\omega/c$, $c$ is the speed of light in vacuum, $\delta$ is the dimensionless detuning from resonance rescaled to the 
dephasing rate, $E_S$ is the saturation field and ${\bf E}$ is the electric field of the optical wave. For weak optical 
fields much smaller than the saturation field, the full-saturated susceptibility can be approximated by its first-order 
Taylor expansion in terms of $|{\bf E}/E_S|^2$, where the zeroth order term 
$\epsilon_b + \alpha (\delta - i)/(1 + \delta^2)$ accounts for linear susceptibility and gain, while the first order 
term $\alpha (i-\delta)/(1 + \delta^2)^2 |{\bf E}/E_S|^2$ accounts for focusing/defocusing nonlinearity (depending on 
the sign of the detuning $\delta$) and nonlinear saturation of the gain. 

Optical propagation in the amplifying plasmonic array sketched in Fig. \ref{Fig1} can be modeled by the following pair of 
coupled mode equations (CMEs) \cite{MariniOL2010}:
\begin{eqnarray}
&& i \frac{ d L_n}{ d z} - i \eta L_n + \kappa ( R_n - R_{n-1} ) + \gamma | L_n |^2 L_n = 0 , \label{CMEq1} \\
&& i \frac{ d R_n}{ d z} - i \eta R_n + \kappa ( L_n - L_{n+1} ) + \gamma | R_n |^2 R_n = 0 , \label{CMEq2}
\end{eqnarray}
where $\eta = \eta' + i \eta''$, $\eta'>0$ is the effective gain parameter,  $\eta''$ is the linear phase shift induced by 
two-level atoms ($\eta'' = 0$ at resonance), $\kappa,\gamma$ are the coupling and nonlinear coefficients and $L_n,R_n$
are the left and right dimensionless field amplitudes at every $n$-th dielectric slot (see Fig. \ref{Fig1}). The longitudinal 
coordinate $z$ is normalized to the scaling length $z_0$, which is arbitrary and can be chosen conveniently. In what follows, 
we will set $z_0$ to be the coupling  length, so that $\kappa = 1$ and $\eta,\gamma$ are complex dimensionless constants. At 
optical frequencies, assuming a coupling length of the order of $z_0 \simeq 1 \mu m$, realistic values for the gain parameter 
are of the order $|\eta| \simeq 10^{-2}$, since amplification of SPPs has been experimentally demonstrated over a distance 
$d \simeq 100 \mu m$ \cite{BeriniNatPhot2011}. The full field ${\bf E}$ is given by the linear superposition 
${\bf E}({\bf r},t) = E_S \sum_{n=-\infty}^{+\infty} \left\{ L_n {\bf e}_{L,n}({\bf r}_{\bot}) + 
R_n {\bf e}_{R,n}({\bf r}_{\bot}) \right\} e^{i \beta z - i \omega t}$, where ${\bf r}_{\bot} = (x , y)$, the (dimensionless) 
vectors ${\bf e}_{L,n}({\bf r}_{\bot})$, ${\bf e}_{R,n}({\bf r}_{\bot})$ are the unperturbed linear mode profiles and $\beta$ 
is the propagation constant of SPPs at every metal-dielectric interface. A full detailed derivation of 
Eqs. (\ref{CMEq1},\ref{CMEq2}) and analytical expressions for the coefficients $\eta,\kappa,\gamma$ are given in Refs. \cite{MariniOL2010,MariniPRA2010,SkryabinJOSAB2011}. Note that the following calculations are not dependent on the 
particular value of the saturation field $E_S$, which scales the optical field. Owing to the dual chirality of alternating 
metal-dielectric interfaces (metal-dielectric and dielectric-metal), the system is inherently binary and every SPP is coupled 
with left and right adjacent SPPs by means of two different coupling coefficients $\kappa_L$, $\kappa_R$. However, it is 
possible to adjust the width of the dielectric slabs ($w_d$) and metallic stripes ($w_m$) in order to achieve the condition 
$\kappa_L = - \kappa_R = \kappa$ \cite{MariniOL2010}. The nonlinear coefficient is complex $\gamma = \gamma' + i \gamma''$, 
the real part can be either positive or negative depending on the sign of the detuning $\gamma' \propto \delta/(1+\delta^2)^2$, 
while the imaginary part is always positive $\gamma''>0$ and accounts for the nonlinear saturation of gain. Note that 
Eqs. (\ref{CMEq1},\ref{CMEq2}) are invariant under reflection in the $x$-direction 
($n \rightarrow - n$, $L_n \rightarrow R_{-n}$, $R_n \rightarrow L_{-n}$), due to the inherent chiral symmetry of the total system.

\begin{figure}[t]
\centering
\begin{center}
\includegraphics[width=0.225\textwidth]{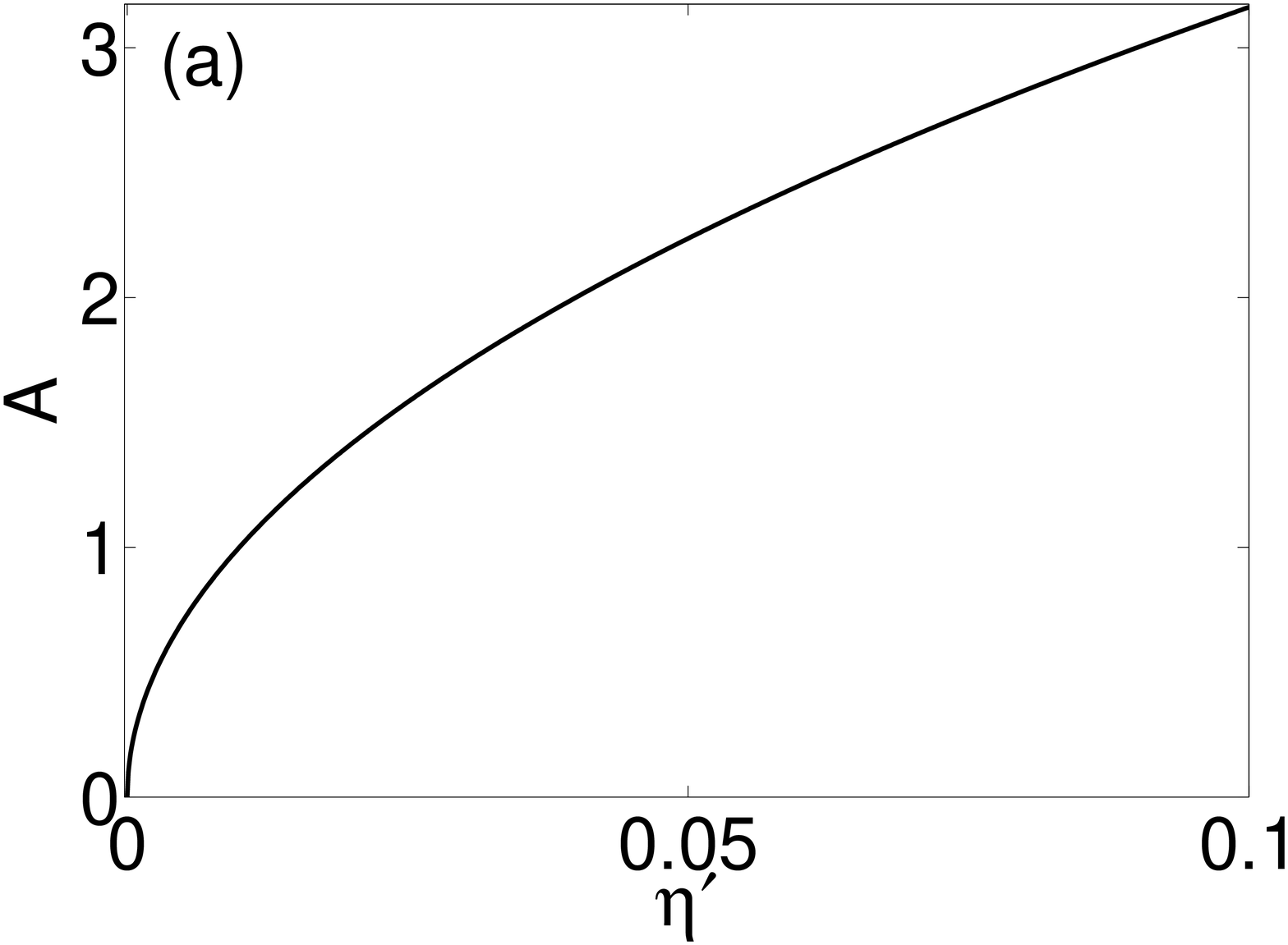}
\includegraphics[width=0.225\textwidth]{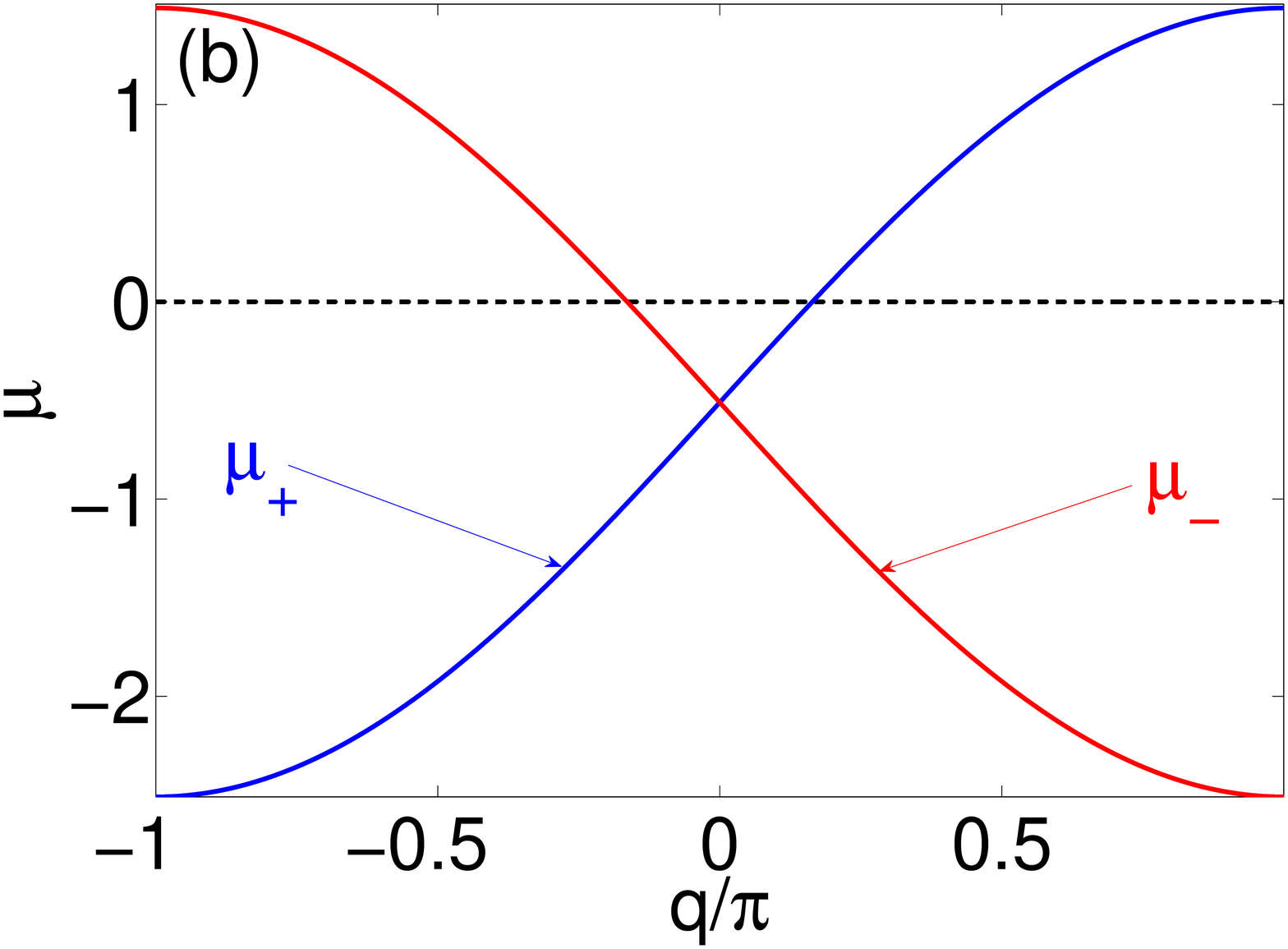}
\caption{(a) Amplitude of the dissipative nonlinear mode $A$ as a function of the effective gain $\eta'$ for $\gamma'' = 0.01 i$. 
(b) Nonlinear dispersion $\mu_{\pm}(q)$ as a function of $q/\pi$ for $\eta = 0.01 - 0.5 i$, $k=1$ and $\gamma = -0.01 + 0.01 i$. 
Blue and red curves represent the two dispersion branches $\mu_+,\mu_-$. The black dashed line denotes $\mu = 0$.}
\label{Fig2}
\end{center}
\end{figure}

\paragraph{Vacuum expectation value --} Owing to the externally pumped active inclusions, small perturbations of the 
vacuum state $L_n = R_n = 0$ are exponentially amplified at a rate $\eta'$. Instability develops until nonlinear effects 
become important and nonlinear gain saturation comes into play counterbalancing the linear amplification. Homogeneous 
nonlinear stationary modes of Eqs. (\ref{CMEq1},\ref{CMEq2}) can be found by taking the Ansatz 
$L_n = L_0 e^{i q n + i \mu z}$, $R_n = R_0 e^{i q n + i \mu z}$, where $q$ is the transverse momentum and $\mu$ is the 
nonlinear correction to the unperturbed propagation constant $\beta$. As a consequence of the dissipative nature of the 
system, the amplitudes $L_0,R_0$ do not remain arbitrary and their moduli are fixed to be $A = \sqrt{\eta'/\gamma''}$. 
The nonlinear correction to the propagation constant is given by $\mu_{\pm} = \eta'' \pm 2 k \sin(q/2) + \gamma' \eta ' / \gamma''$. 
The amplitude of the dissipative nonlinear mode $A$ is plotted as a function of the effective gain 
parameter $\eta'$ in Fig. \ref{Fig2}a, while the nonlinear dispersion $\mu_{\pm}(q)$ is depicted in Fig. \ref{Fig2}b. 
Note that, due to the inherent alternate coupling of the system, the nonlinear dispersion is characterized by a Dirac 
diabolical point at $q=0$ \cite{NamOE2010}. At this special point, the phases of both amplitudes $L_0,R_0$ remain arbitrary. 
Conversely, for $q \neq 0$ the mode amplitudes are fixed to $R_0 = \mp i e^{i q / 2} L_0$ and only a global phase is left 
arbitrary.

\begin{figure}[t]
\centering
\begin{center}
\includegraphics[width=0.45\textwidth]{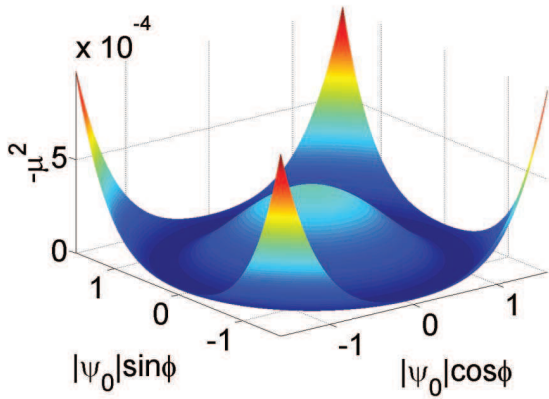}
\caption{Optical analogue of the mexican hat potential describing spontaneous symmetry breaking in quantum field theory:
$-\mu^2$ is plotted against $|\psi_0|cos\phi$, $|\psi_0|sin\phi$, where $|\psi_0|$ is the field amplitude and $\phi$ is the
relative phase between the spinor components. The plot is made by taking the parameters $\eta' = 0.02$, $\kappa = 1$, 
$\gamma''=0.01i$.}
\label{Fig3}
\end{center}
\end{figure}

\paragraph{Nonlinear Dirac-like equation --} As mentioned in the introduction, BWAs have been used to mimic phenomena in both 
non-relativistic and relativistic quantum mechanics \cite{LonghiLasPhotRev2009,DreisowPRL2010}, since CMEs can be converted 
into the one-dimensional relativistic Dirac equation \cite{note}. Defining the two-component spinor $\psi = [L_n(z) , R_n(z)]^T$, 
if the transversal patterns of the amplitudes $L_n$, $R_n$ are smooth, one can take the continuous limit by
introducing the continuous spatial coordinate $n \rightarrow x$. In this limit, the spinor satisfies the (1+1)D nonlinear 
Dirac-like equation
\begin{equation}
i \partial_z \psi - i \eta \psi + i \kappa \hat{\sigma}_y \partial_x \psi + \gamma G ( \psi ) = 0, \label{NLDiracEq}
\end{equation}
where $G ( \psi ) = ( |L|^2 L , |R|^2 R )^T$ is the nonlinear spinorial term and $\hat{\sigma}_y$ is the $y$-Pauli matrix.
In what follows, we will focus on the case where the angular frequency of SPPs $\omega$ coincides with the two-level atom
resonant frequency and thus the detuning $\delta$ vanishes: $\delta = 0$, so that $\eta = \eta'$ and $\gamma = i \gamma''$. 
In this case Eq. (\ref{NLDiracEq}) is analogous to the (1+1)D Thirring model \cite{ThirringAnnOfPhys1958} with imaginary mass 
and nonlinear terms, describing the dynamics of fermionic tachyons. Optical analogues of fermionic tachyons have been 
recently investigated in optical graphene and in topological insulators \cite{SzameitPRA2011,ApalkovEPL2012}. Note that 
Eq. (\ref{NLDiracEq}) is a {\it Dirac-like} equation, since the ``mass term'' ($- i \eta \psi$) is different from previously
studied standard formulations \cite{DreisowPRL2010,SzameitPRA2011}, and is responsible for the existence of unstable 
tachyon-like particles. Owing to amplification, vacuum dynamically acquires a stable expectation value and the ensuing final 
state is the optical analogue of a condensate of stable fermionic particles with non-negative squared mass. In turn, this process 
is commonly named {\it tachyon condensation}, e.g. in the context of open string field theories \cite{HellermanJHEP2013}.

\paragraph{Spontaneous symmetry breaking --} Note that, analogously to the Thirring \cite{ThirringAnnOfPhys1958}, 
sine-Gordon \cite{ColemanAmmPhys1976} and Nambu-Jona-Lasinio \cite{{NambuPhysRev1960}} models, Eq. (\ref{NLDiracEq}) is 
chirally symmetric since it is left invariant under reflection $x\rightarrow -x$ if the spinor components are transformed 
as $L(x)\rightarrow R(-x)$, $R(x)\rightarrow L(-x)$. In turn, while the unstable vacuum state $\psi = 0$ is chirally 
symmetric, the nonlinear homogeneous mode $\psi = \psi_0 e^{i \mu z}$ with finite amplitude $ \psi_0$ and propagation
constant $\mu$ [where $\mu^2 = -(\eta' - \gamma''|\psi_0|^2)^2$] breaks the chiral symmetry. The optical analogue of energy 
is represented by the propagation constant $\mu$ and the system spontaneously evolves to states where $-\mu^2$ is minimum.
In Fig. \ref{Fig3}, we plot $-\mu^2 = (\eta' - \gamma''|\psi_0|^2)^2$ as a function of the mode amplitude $|\psi_0|$ and 
the relative phase between the spinor components $\phi$. We find the characteristic mexican hat profile, which constitutes 
the archetypical potential describing spontaneous symmetry breaking in QFT. Our optical analogue of tachyon condensation 
thus drives the physical system to a stable state with broken chiral symmetry where $-\mu^2$ is minimum and particles 
(i.e. optical states) with non-negative squared mass are generated.

\begin{figure}[t]
\centering
\begin{center}
\includegraphics[width=0.225\textwidth]{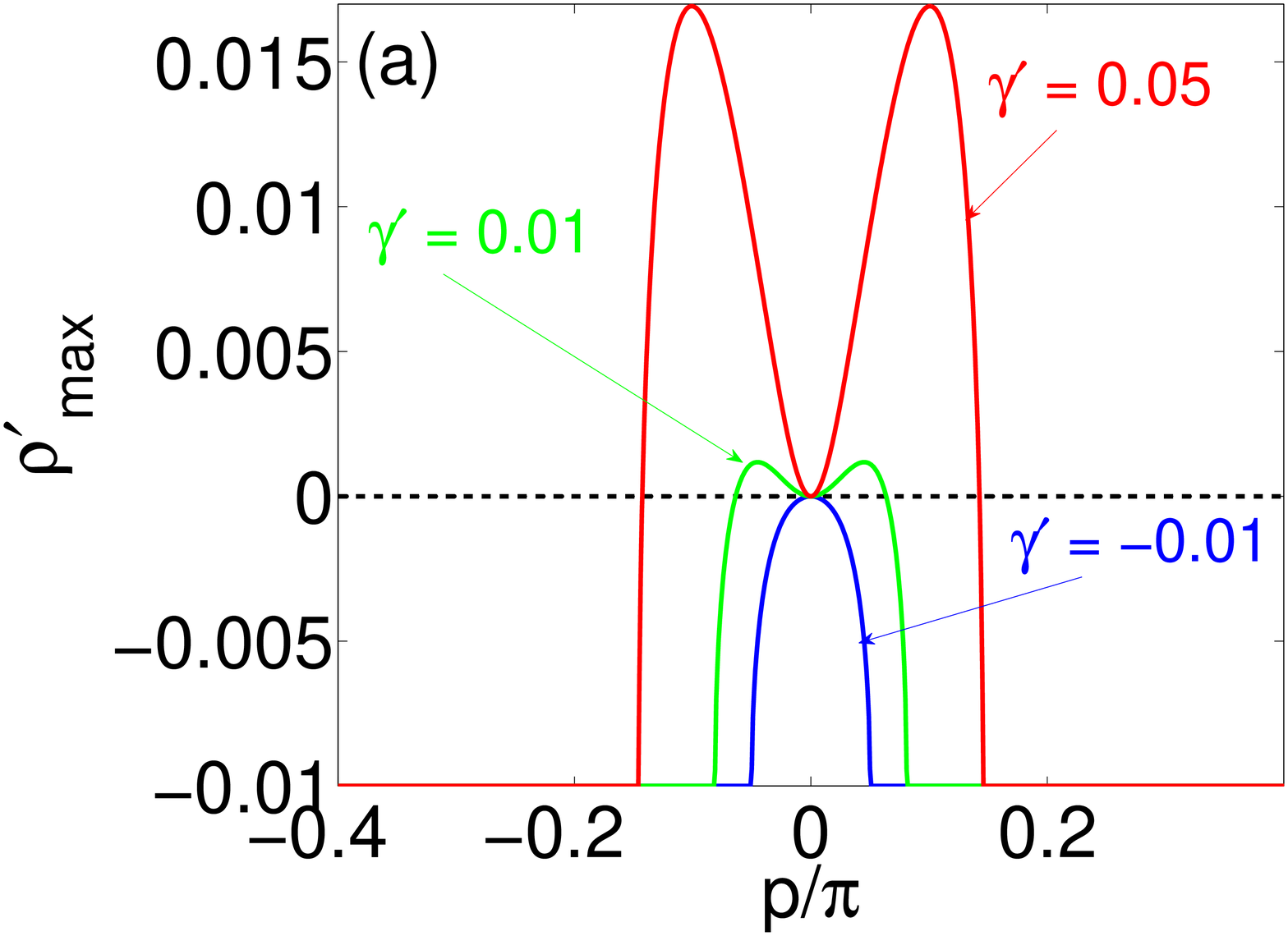}
\includegraphics[width=0.225\textwidth]{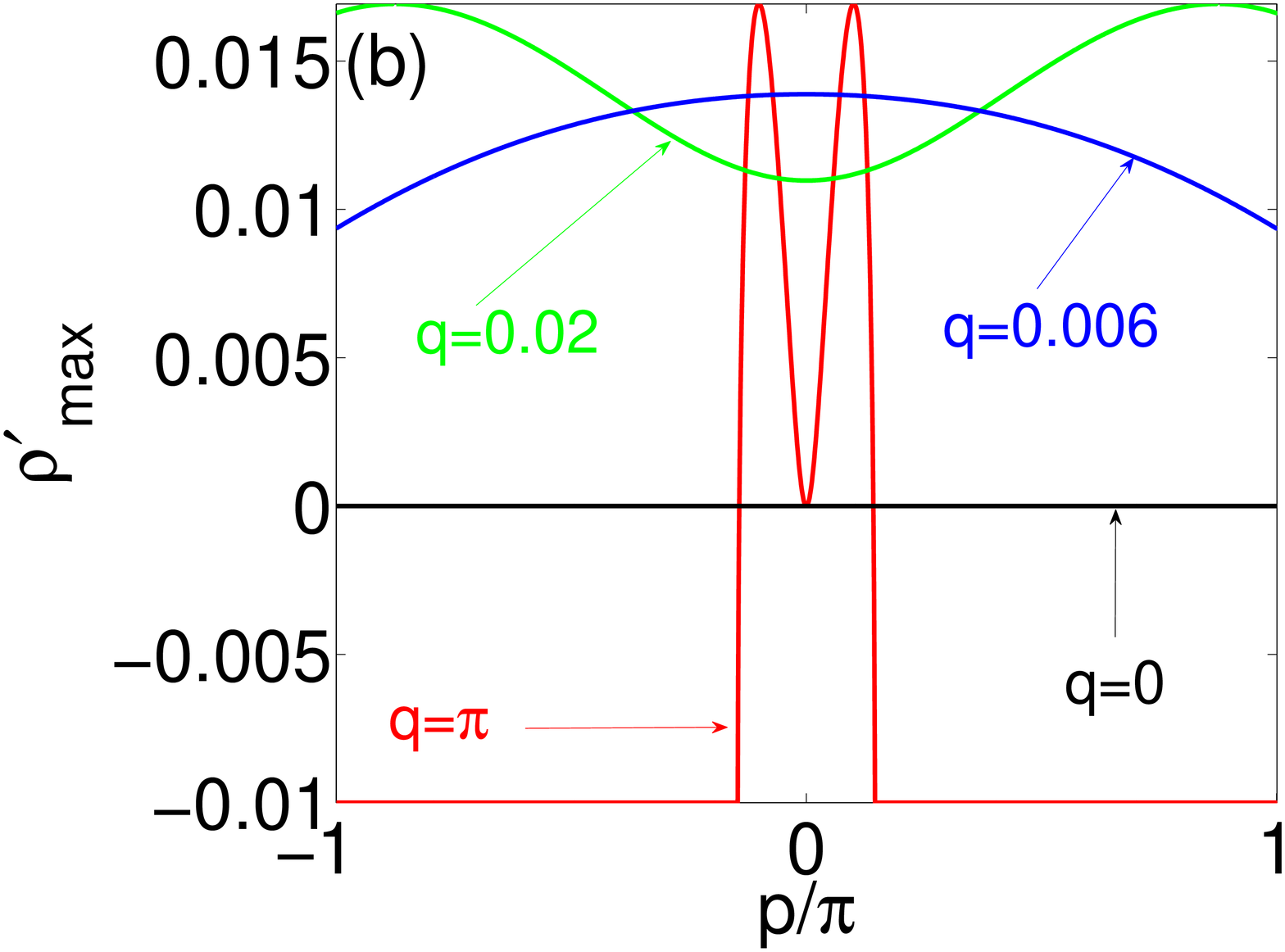}
\caption{Maximum instability eigenvalue $\rho'_{max}$ as a function of the transverse momentum $p$ of perturbating waves. 
(a) $\rho'_{max}$ vs $p$ for $q = \pi$, $\kappa = 1$, $\eta' = 0.01$, $\gamma'' = 0.02i$ and several values of $\gamma'$.
Blue, green and red curves correspond to $\gamma' = - 0.01,0.01,0.05$. The black dashed line denotes the instability threshold
$\rho'_{max} = 0$.
(b) $\rho'_{max}$ vs $p$ for $\kappa = 1$, $\eta' = 0.01$, $\gamma = 0.05 + 0.02i$ and several values of $q$.
Black, blue, green and red curves correspond to $q = 0, 0.006, 0.02, \pi$. }
\label{Fig4}
\end{center}
\end{figure}

\paragraph{Modulational instability --} In order to study the stability of the nonlinear homogeneous stationary mode with
broken chiral symmetry under transversal modulation, we perturb it with small amplitude waves carrying transverse momentum $p$:
\begin{eqnarray}
&& L_n =  \left[ L_0 + l_+ e^{i p n + \rho z} + l_-^* e^{ - i p n + \rho^* z} \right] e^{ iqn + i \mu z} , \label{MIEqL} \\
&& R_n =  \left[ R_0 + r_+ e^{i p n + \rho z} + r_-^* e^{ - i p n + \rho^* z} \right] e^{ iqn + i \mu z} . \label{MIEqR}
\end{eqnarray}
Inserting Eqs. (\ref{MIEqL},\ref{MIEqR}) into Eqs. (\ref{CMEq1},\ref{CMEq2}) and linearizing for small $|l_{\pm}|$, 
$|r_{\pm}|$ one finds a fourth order homogeneous system of algebraic equations 
$[ \hat{\cal M} - \rho \hat{1} ] {\bf v} = 0$, where 
${\bf v} = (l_+ , l_- , r_+ , r_-)^T$, $\hat{1}$ is the identity matrix, $\hat{\cal M}$ is the linearized system matrix and 
$\rho_1$, $\rho_2$, $\rho_3$, $\rho_4$ are the instability eigenvalues of $\hat{\cal M}$ that we have calculated 
numerically. Instability occurs if one of the complex eigenvalues $\rho$ has a positive real part. In Fig. \ref{Fig4}a,b
we plot the real part $\rho_{max}^{'}$ of the most unstable eigenvalue as a function of the transverse momentum $p$ of perturbating waves 
for $\kappa = 1$, $\eta' = 0.01$, $\gamma'' = 0.02i$. In Fig. \ref{Fig4}a, $\rho'_{max}$ is plotted for $q = \pi$ and 
$\gamma'=-0.01,0.01,0.05$ (blue, green and red curves), while in Fig. \ref{Fig4}b the real part of the nonlinear 
coefficient is fixed to $\gamma' = 0.05$ and $q = 0, 0.006, 0.02, \pi$ (black, blue, green and red curves). In both figures,
the instability eigenvalues were calculated for the modes of the upper branch ($\mu_+$, see Fig. \ref{Fig2}). Note that, for 
$q \neq 0$, stability depends on the sign of $\gamma'$ (instability for $\gamma'>0$ and stability for $\gamma'<0$) and thus 
on the sign of the detuning $\delta$. Conversely, at the Dirac point $q=0$, homogeneous nonlinear waves are always 
marginally stable ($\rho'_{max} = 0$). 

\begin{figure}[t]
\centering
\begin{center}
\includegraphics[width=0.225\textwidth]{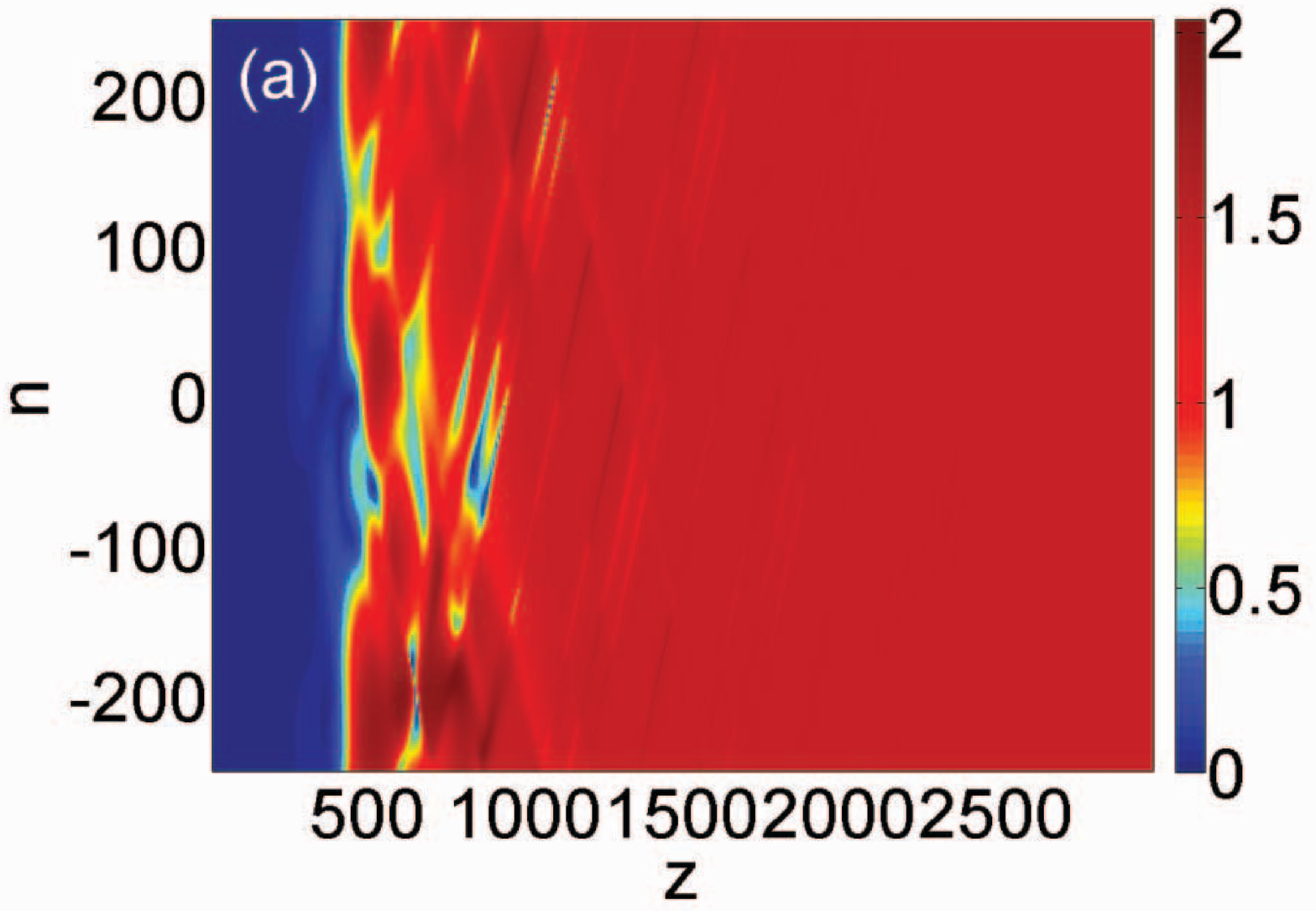}
\includegraphics[width=0.225\textwidth]{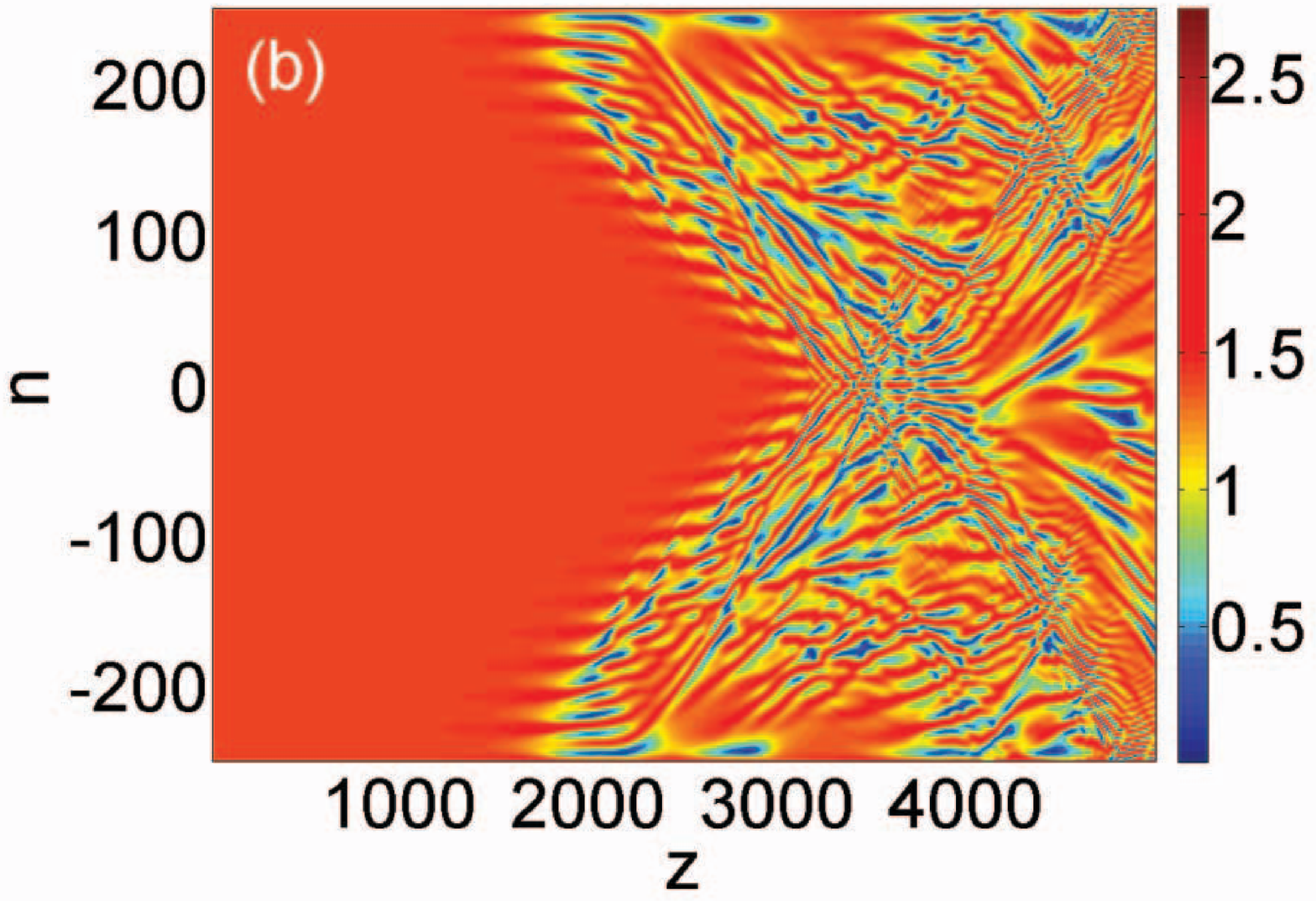}
\includegraphics[width=0.225\textwidth]{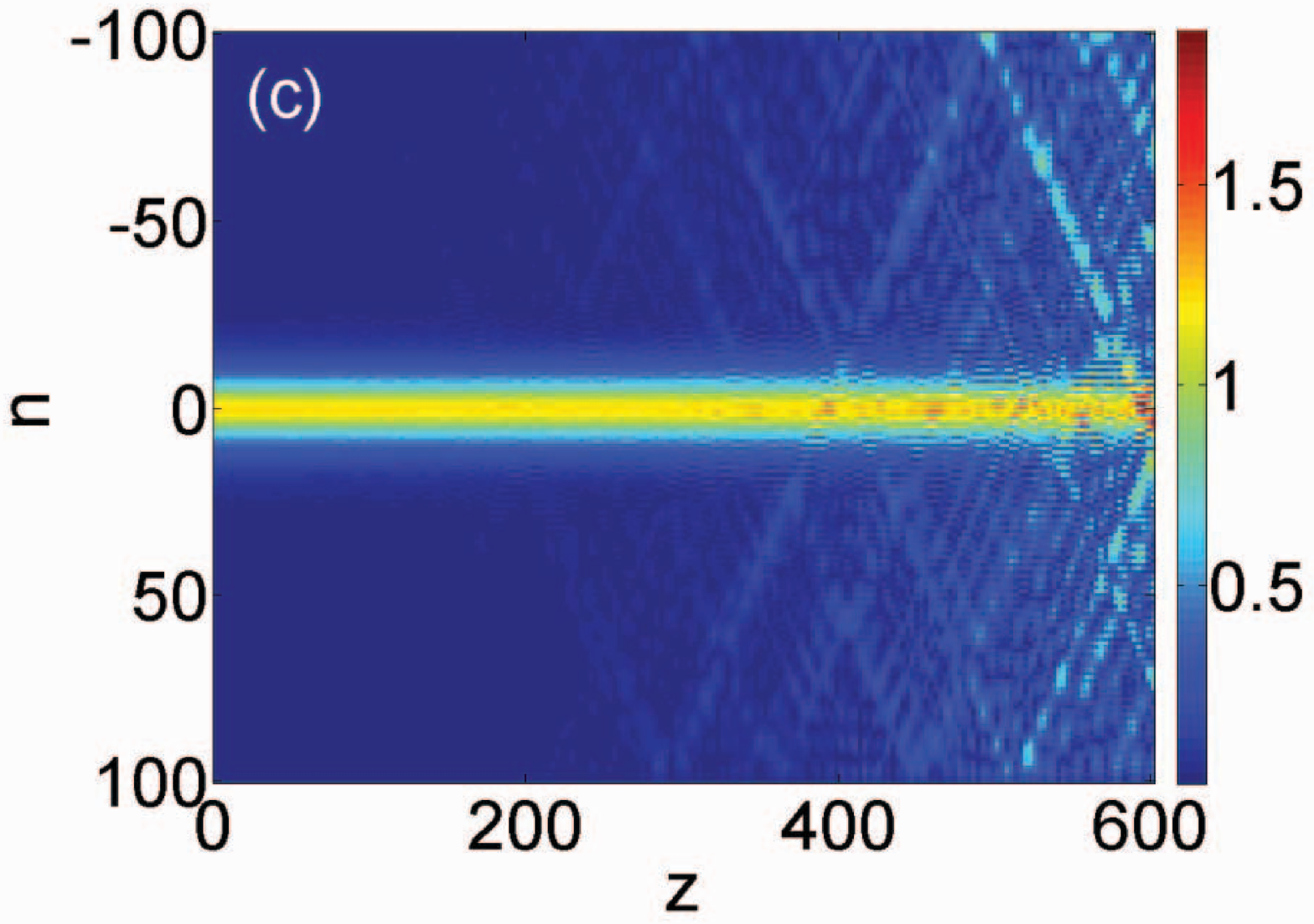}
\includegraphics[width=0.225\textwidth]{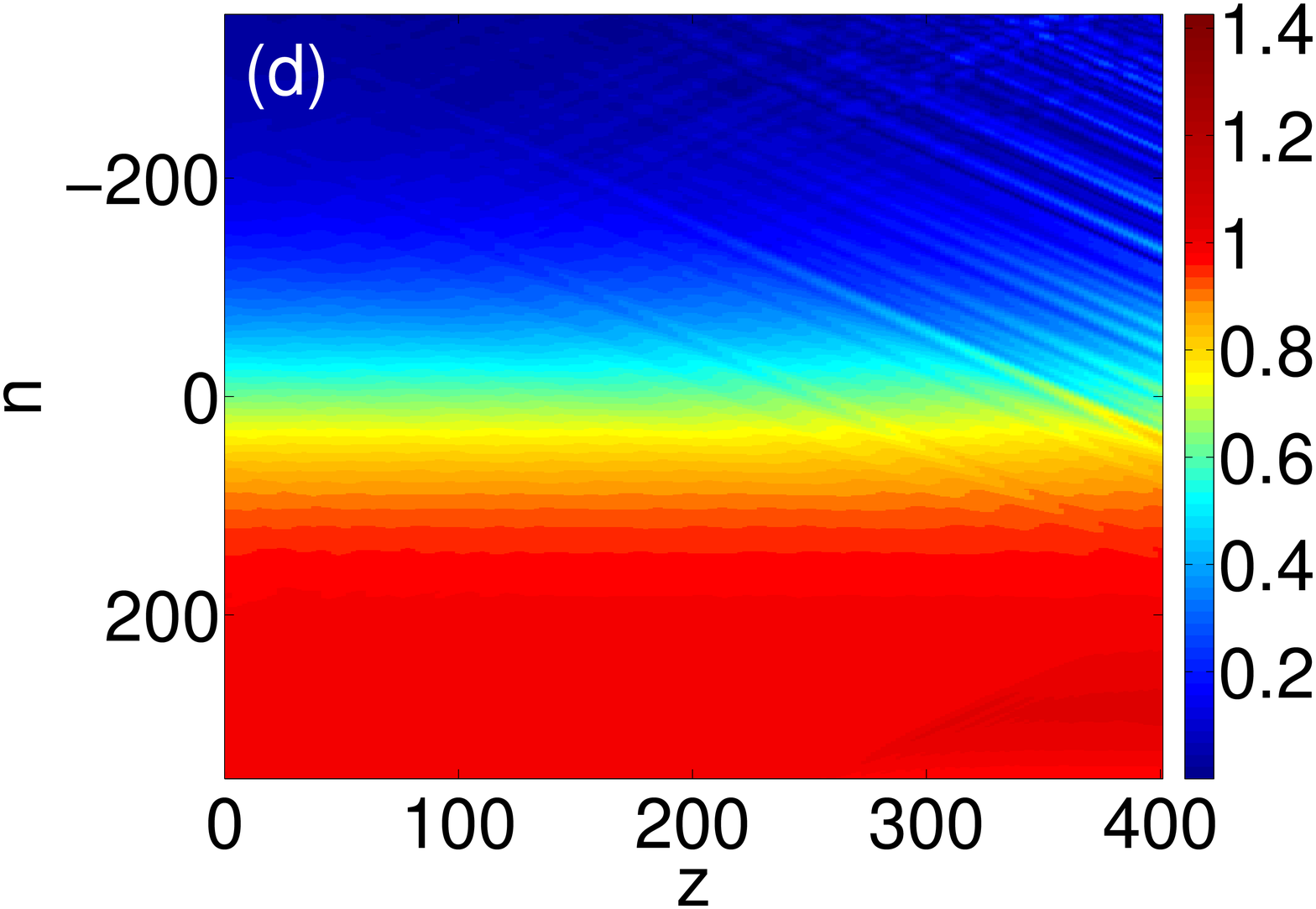}
\caption{Propagation contour plots of the left field amplitude $|L_n|$ for several input conditions $L_n (0)$, $R_n(0)$
weakly perturbed with random noise:
(a) vacuum state $L_n(0) = R_n(0) =  0$, (b) nonlinear homogeneous mode at the band edge 
$L_n(0) = R_n(0) = \sqrt{\eta'/\gamma''}e^{i \pi n}$, (c) bright soliton with $q = \pi$ and (d) kink soliton with $q=0$.  
Numerical integration is taken with the parameters $\eta = 0.01$, $\kappa = 1$ and (a) $\gamma = 0.01 i$, 
and (b,c,d) $\gamma = 0.01 + 0.01 i$.
}
\label{Fig5}
\end{center}
\end{figure}

These predictions have been also confirmed by the direct numerical integration of 
Eqs. (\ref{CMEq1},\ref{CMEq2}) using a fourth order Runge-Kutta algorithm. In the panels of Fig. \ref{Fig5}, we contour 
plot the modulus of the left optical field $|L_n|$ as a function of the SPP index $n$ and of the propagation direction 
$z$ for different input conditions. In Fig. \ref{Fig5}a, we set as initial condition a small random perturbation of the 
vacuum state, which is unstable and dynamically converges to the stable nonlinear homogeneous mode at the Dirac point $q=0$,
which represents the vacuum expectation value. In Fig. \ref{Fig5}b, we perturb the homogeneous nonlinear mode of the upper branch 
at the band edge $q = \pi$ with small random perturbations, finding a modulationally unstable chaotic dynamics. Indeed, 
modulational instability is strongly related to the presence of topological defects, which we have found in the present 
system as kink, bright and dark dissipative solitons. Due to the instability of the vacuum background, topological defects 
are also unstable and behave as strange attractors for the dynamical system. We have numerically calculated the bright and 
kink soliton profiles by using the shooting method. In Figs. \ref{Fig5}c,d, we perturb bright and kink solitons with small 
random waves finding that the background noise is amplified and eventually destroys the solitons. 

\paragraph{Conclusions --} In this Letter we have studied an optical analogue of spontaneous symmetry breaking induced
by tachyon condensation. We focused our attention on amplified SPPs propagating at every interface of a metal-dielectric 
stack, but our results are valid also for amplifying BWAs with alternating coupling. Optical propagation is modeled 
through CMEs, which in the continuous limit converge to a nonlinear Dirac-like equation that conserves chirality. 
We find that the vacuum is unstable and the system spontaneously evolves to a stable homogeneous state with broken chiral 
symmetry. This symmetry breaking is accompanied by the formation of propagating optical modes, which correspond to particles 
with non-negative squared mass in the QFT/optics analogy. We studied the modulational instability of the nonlinear modes 
of the system, and found that at the Dirac point instability never occurs. This paves the way for using amplifying 
plasmonic arrays as a classical laboratory for spontaneous symmetry breaking effects in quantum field theory. We also envisage 
that further investigations and developments of QFT/optical analogies may be found in the context of nonlinear ${\cal PT}$-symmetric 
optical systems.   

AM, SR, FB (Max Planck Research Group) and TXT (Max Planck Partner Group) are supported by the German Max Planck Society for 
the Advancement of Science (MPG).


\begin{thebibliography}{10}
\bibitem{note} It is important to note that, although the linear Dirac equation is Lorentz invariant, 
its {\em nonlinear} version in general is not. However, here this issue is not a concern, since in the paraxial 
approximation Lorentz invariance is not a requirement for a physically meaningful model.
\end{thebibliography}
\end{document}